\documentclass[preprint,aps,superscriptaddress]{revtex4}
\usepackage{graphicx}
\usepackage{amsmath}
\usepackage{amssymb}
\newcommand{\be}{\begin{equation}}
\newcommand{\ee}{\end{equation}}
\newcommand{\bea}{\begin{eqnarray}}
\newcommand{\eea}{\end{eqnarray}}
\newcommand{\pr}{\partial}
\newcommand{\nno}{\nonumber}
\newcommand{\bse}{\begin{subequations}}
\newcommand{\ese}{\end{subequations}}

\begin{document}
\title{Semi-realistic Bouncing Domain Wall Cosmology}
\author{Debaprasad Maity
\footnote{debu.imsc@gmail.com}}
\affiliation{Department of Physics and Center for
Theoretical Sciences, National Taiwan
University, Taipei 10617, Taiwan}
\affiliation{Leung Center for Cosmology and Particle Astrophysics\\
National Taiwan University, Taipei 106, Taiwan}

\begin{abstract}
In this paper we constructed a semi-realistic  
cosmological model in a dynamic domain wall framework. 
Our universe is considered to be 
a (3+1) dimensional dynamic domain wall  
in a higher dimensional Einstein-Maxwell-Born-Infeld 
background. One of our interesting
outcomes from the effective Hubble equation for the domain wall
dynamics is that it contains an additional
component of "dark matter" which is induced 
from the charge of the bulk Born-Infeld gauge field.
In this background spacetime we have studied the cosmological 
dynamics of the domain wall. 
In addition to the Born-Infield gauge field if we consider additional 
pure gauge field, a non-singular bounce happens at the early stage 
with a smooth transition between contracting and expanding phase.

\bigskip
\noindent
PACS number: 04.50.-h, 04.60.Cf, 04.20.Jb
\end{abstract}
\maketitle
\section{Introduction}\label{intro}
Standard model of cosmology has already been proved to be one of the 
most successful models in physics. In spite of its success in accounting
various cosmological as well astrophysical observations, the model 
is plagued with some basic fundamental problems. 
On of those is the famous big-bang singularity problem. 
In the standard big-bang model if one goes backward in time, it
hits the singularity at finite time. 
Many different approaches have been proposed over the years to avoid 
this problem. One of the approaches that has gained considerable interest
is in the framework of braneworld. In this approach our universe 
is identified with a four dimensional hyper-surface [1-9]
moving in the extra dimensional spacetime. A co-dimension one hyper-surface
is technically  called domain wall. Through out our paper we will
consider the dynamics of a domain wall. In this framework 
it has been shown that dynamics of a domain wall in the extra dimension 
mimics usual Hubble equation of standard cosmology with the additional
components of induced invisible energy. This gives 
us a possibility of studying the cosmology in a new perspective
\cite{kraus,csaki}. 
One of the important aspects of this framework is that
Hubble equation of motion for the domain wall emerges from the
boundary condition across its position in the extra dimension which is 
known as Israel junction condition\cite{israel}. 
Furthermore different parameters of the 
bulk spacetime solution effectively act as a source of 
invisible energy density with different equations of 
state on the domain wall. By tuning those parameters in a 
model under consideration, one can in principle construct
viable cosmologies with a bounce which avoids the usual big-bang 
singularity. Furthermore, it is an interesting point to note that
by tuning those bulk parameters one can also construct a model universe
with an induced "dark radiation" and "dark matter" component in addition to 
the bounce with a transition from 
contracting phase followed by the standard expanding phase of the universe 
\cite{sudipto,novello}.
Motivated by our previous study, in his paper we constructed 
such a semi-realistic bouncing domain wall cosmological model 
without introducing standard dark matter component on the domain wall 
\cite{debu}.

As a follow up of our previous study we will 
construct a simple cosmological model of 
dynamic domain walls in the background of 
Maxwell and Born-Infield gauge filed along the line of \cite{chamblin}.
Let us mention at this point that we consider two types 
of gauge fields. One corresponds to the standard Maxwell field ${\cal A}_A$
and other one is Born-Infeld gauge field ${\cal B}_B$. 
Purpose of taking those two different types of gauge field will be 
apparent as we proceed.
Motivation to consider both kind of gauge fields could be coming
from string theory. Born-Infeld type higher derivative action
naturally arises in string theories in their low-energy 
effective action. In addition to the 
the gauge field the effective action also contains
an infinite series of higher curvature terms in the gravity sector. 
For our present purpose in this report, we will ignore 
those higher spacetime curvature terms. 
For simplicity, in this paper we consider the gauge field higher
derivative terms like Born-Infeld gauge field. 
We have solved analytically the equations
of motion with the appropriate junction condition at the position
of the domain wall. There exists three different types of solution depending
upon the choice of parameters. We have already discussed in details
about part of those solutions in our previous works \cite{debu}.
For the present purpose, we have chosen the simplest but phenomenologically
appealing solution which we find has an interesting cosmological
implications with regard to our aforementioned motivation to construct
a domain wall cosmology.

We structured this report as follows: In section \ref{sec1}, we
will start with a generic action corresponding to a domain wall
moving in Maxwell-Born-Infield-dilaton background. 
In order our paper to be self contained, we will give the 
general analysis with the dilaton field in this section.  
In the subsequent section \ref{sec2} as we mentioned before 
we will consider a particular bulk background with 
a trivial dilaton configuration. We take the static bulk metric ansatz and
study the dynamics of the domain wall in this static background.
We get semi-realistic bouncing domain wall cosmology with 
dark radiation and dark matter like energy components induced from
the bulk black hole charges. 
In section \ref{sec3}, 
we consider more realistic case where we have matter
field localized on the brane. This has changed the effective Hubble
equation significantly. We find the corresponding constraints on the
bulk spacetime parameters so that we have a non-singular bouncing cosmology
even with the standard matter field.
We also discussed about the possible 
constraints on the parameters of our solution from the cosmological 
observations.
In section \ref{sec4} we will discuss about the perturbation equations
across the domain wall.
Finally, in section \ref{con}, we do some concluding remarks and 
describe some futures directions to work.

\section{Einstein equations and Boundary conditions} \label{sec1} 
We start with a general action of the Einstein-Maxwell-Born-Infeld-dilaton
system in an arbitrary spacetime dimension $n$. 
The action takes the from 
\begin{equation} \label{action}
S = \int d^n x\sqrt{-g}\left( \frac 1 2 R ~-~ \frac 1 2 
\pr_{A} \phi \pr^{A} \phi ~-~ V(\phi) ~-~ \frac {1} 4 e^{- 2 \zeta \phi} 
F_{AB}F^{AB}~+~ {\cal L}(G,\phi)\right) ~+~ S_{DW},
\end{equation}
where action for the domain wall is
\begin{center}
$S_{DW} ~=~  ~-~ \int d^{n-1} x \sqrt{-h} \left(
\{K\} + \bar{V}(\phi) \right)$.
\end{center}
The expression for ${\cal L}(G,\phi)$  is 
\begin{equation}
{ L(G,\phi)} = {4{\lambda}^2 e^{2 \gamma \phi} \Big 
(1-\sqrt{1+\frac{e^{-4 \gamma \phi} G^{AB} 
G_{AB}}{2{\lambda}^2}}\Big)},
\label{}
\end{equation}
where $\lambda$ is a constant parameter with the dimension of mass. 
$G_{AB}=\pr_A {\cal B}_{B}-\pr_B {\cal B}_A$ is the Born-Infeld field 
strength and $F_{BD}=\pr_B {\cal A}_D - \pr_{D} {\cal A}_B$ 
is the field strength of the Maxwell field  $A_D$.
$h$ is the determinant of the induced metric $h_{AB}$ on the domain wall.
$K$ is the trace of the
extrinsic curvature $K_{ab}$ of the domain wall.

Corresponding Einstein equations turn out to be
\bea
&&R_{AB} = T_{AB}^{\phi} + T_{AB}^{{\cal A}} + T_{AB}^{\cal B} ~~~\\
&&D_C \pr^C \phi - \frac {\pr (\phi)}{\pr \phi} + 8 \lambda^2 \gamma  e^{2 \gamma \phi}\left\{2 {\cal Y} 
\frac {\pr {\cal L}}{\pr {\cal Y}} - {\cal Y} \right\} + \frac 1 2 \zeta
 e^{- 2 \zeta \phi}F_{AB}F^{AB} = 0 \\
&& D_A\left( e^{- 2 \gamma \phi} \frac {\pr {\cal L}}{\pr {\cal Y}} G^{AB}\right) = 0\\ 
&& D_A\left (e^{- 2 \zeta \phi} F^{AB}\right) = 0
\eea
where, various energy momentum tensor components are 
\bea
T_{AB}^{\phi} &=& \pr_{A} \phi \pr^A \phi + \frac 2 {n -2} V(\phi) g_{AB} 
~~;~~T_{AB}^{\cal A} = \frac 1 2 e^{- 2 \zeta \phi} \left( 2 F_{AC}F_B^C - 
\frac {1}{n-2}
 F_{CD}F^{CD} g_{AB}\right) \nno\\
T_{AB}^{\cal B} &=&  \frac {8 \lambda^2} {(n-2)}
e^{2 \gamma \phi}\left\{2 {\cal Y} \frac {\pr {\cal L}}{\pr {\cal Y}} - {\cal Y} \right\} g_{AB}
- 8 e^{- 2 \gamma \phi} \frac {\pr {\cal L}}{\pr {\cal Y}} G_{AC}G_B^C \nno 
\eea
 $D_A$ is co-variant derivative with respect to the bulk 
metric and ${\cal Y} = \frac{e^{-4 \gamma \phi} G^{AB} 
G_{AB}}{2{\lambda}^2}$. In addition to the above equation we need to satisfy
the following Israel junction conditions 
\bea \label{bond}
&&\{K_{MN}\} = - \frac 1 {n -2} \bar V(\phi) h_{MN} \\
&&\{n^{M}\pr_{M} \phi\} = \frac {\pr \bar{V}(\phi)}{\pr \phi}
\eea
where, $n^M$ is the unit normal to the domain wall.
$R$ is the curvature scalar.

In the subsequent analysis we will consider our model 
enjoying reflection symmetry($Z_2$) across the domain wall. 
Considering a static spherically symmetric bulk metric  
\be \label{metric}
ds^2 = - N(r) dt^2 + \frac 1 {N(r)} dr^2 + R(r)^2 d\Omega_{\kappa}^2
\ee
with $ d\Omega_{\kappa}^2$ being a metric on a $(n -2)$ dimensional space
with a constant curvature  
$\tilde {R}_{ij} = k (n - 3) \tilde {g}_{ij}$ with $ k \in \{-1,0,1\}$,
we are interested to study induced cosmological dynamics on the 
domain wall with a Freedman-Robertson-walker metric 
\bea
ds_{wall}^2 = - d\tau^2 + R(\tau)^2 d\Omega_{\kappa}^2.
\eea
$\tau$ is the
domain wall proper time. 
As one can clearly see from the above construction that the 
radial direction along the extra dimension plays the role of 
scale factor of our 
domain wall universe.

By considering the unit normal to be pointing towards $r < r(t)$ region, one 
can find the following equations consistent with the dynamic domain wall 
in the extra dimension 
\bea \label{E1}
R' = C \bar{V}(\phi) .
\eea
Using above equation in the 
boundary condition for the scalar field one gets
\bea \label{E2}
\frac {\pr \phi}{\pr R} = - \frac {n -2} R \frac 1 {\bar V} \frac {\pr \bar V}
{\pr \phi}
\eea

In the above derivation we have used the expression 
for $K_{ij}$ and $K_{00}$.

So, one can solve the above equation for $\phi$ as a 
function of scale factor R without referring to the bulk scalar
field potential. This is in consistent with the 
dynamic domain wall coupled with a bulk scalar field we mentioned before. 
Now we will solve the full equation of motion in consistent with the
above equations.

\section{Bulk solutions and domain wall cosmology } \label{sec2}
In our previous papers \cite{debu} we have already solved for the 
Born-Infield and Maxwell field coupled with dilaton separately.
In this paper, we will solve them togather and then try so 
study its cosmology.

We consider a class of solution for the both Born-Infeld and Maxwell 
field where all the components of $F^{AB}$ and $G^{AB}$ being
zero except $F^{rt}$ and $G^{rt}$ component. The solution looks like
\bea \label{BIsol}
G^{rt} = \frac {2 Q \lambda e^{2 \gamma \phi}}
{\sqrt{4 Q^2 + \lambda^2 R^{2 n -4}}}~~;~~ F^{rt}= \frac { Q^{\dagger}  
e^{2 \zeta \phi}}{2 R^{ n -2}} 
\eea
where, $Q$ and $Q^{\dagger}$ are the integration constant and 
related to the Born-Infield and electromagnetic charge respectively. 
Where Born-Infield and electromagnetic charges can be expressed as follows
\bea
q = \frac 1 {4\pi} \int_{\Sigma_\infty} e^{- 2 \gamma \phi}~ {^*G} = 
\frac {Q \omega_{n-1}}{4 \pi}~~;~~ q^{\dagger} = \frac 1 {4\pi} \int_{\Sigma_\infty} e^{- 2 \zeta \phi}~ {^*F} = 
\frac {Q^{\dagger} \omega_{n-1}}{4 \pi}
\eea
where, ${^*F}_{AB} = \frac 1 {2 \sqrt{-g}} 
\varepsilon^{ABCD} F^{CD}$ and ${^*G}_{AB} = \frac 1 {2 \sqrt{-g}} 
\varepsilon^{ABCD} G^{CD}$. $\Sigma_{\infty}$ is a hyper-surface
at $R \rightarrow \infty$ . $\omega_{n-1}$ is volume of unity $n$ sphere.
  
Using the solution for the Born-Infeld and Maxwell field and 
the ansatz for the metric Eq.\ref{metric}, 
the remaining equations of motion turn out to be
\begin{subequations}
\bea
&&\frac {R''} {R} = - \frac 1 {n-2} \phi'^2\\
&& \frac 1 {2 R^{n-2}} \left\{N \left(  R^{n-2}\right)'\right\}' -
\frac {k(n-3)(n-2)} {2 R^2} = - V - {\cal T}_{22}(R,Q) - 
\frac { {Q^{\dagger}}^2 } {R^{2n -4}}
e^{2 \zeta \phi}\\
&& \frac {n-2}{4 R^{n-2}} \left( N' R^{n-2}\right)' = - V - {\cal T}_{00}(R,Q)+
\frac {(n-3) {Q^{\dagger}}^2 }{2 R^{2n -4}} e^{2 \zeta \phi} \\
&&\frac 1 {R^{n-2}}\left( \phi' N R^{n-2}\right)' = 
\frac {\pr V(\phi)}{\pr \phi} + 8 \lambda^2 \gamma e^{2 \gamma \phi}  
{\cal E}(r,Q) +
\frac { \zeta {Q^{\dagger}}^2}{R^{n-2}} e^{2 \zeta \phi},
\eea\end{subequations}

where
\bea
{\cal T}_{22}(R,Q) = 4 \lambda^2  e^{2 \gamma \phi} {\cal E}(R,Q)~~;~~
{\cal T}_{00}(R,Q) =
4(n-2) \lambda^2  e^{2 \gamma \phi}\left[\frac {{\cal E}(R,Q)}{n-2} + \frac {{\cal G}(R,Q)}2 \right]
\eea
and
\bea
{\cal E}(R,Q) = \frac {\sqrt{4 Q^2 + \lambda^2 R^{2 n -4}}}{\lambda R^{n -2}} - 1~~~;~~~
{\cal G}(R,Q) = - \frac {4 Q^2}{\sqrt{4 Q^2 + \lambda^2 R^{2 n -4}}} \frac 1 {\lambda R^{n -2}}.
\eea

${\cal T}_{00}$ and ${\cal T}_{22}$ are $tt$  and $xx$ 
components of the energy-momentum tensor for the Born-Infield Lagrangian respectively.

In order to solve, we choose the following Liouville
type brane potential 
\be
\bar V(\phi) = {\bar V}_0 e^{\alpha \phi},
\ee
which provides a straight forward solution for the scalar field $\phi$
and the scale factor $R$ without any specific 
form of the the bulk potential.
\begin{subequations} \label{sol}
\bea 
&&\phi = \phi_0 - \frac {\alpha (n-2)}{\alpha^2(n-2) + 1} log (r)\\
&&R(r) = C {\bar V}_0 e^{\alpha \phi_0} 
r^{\frac 1 {\alpha^2(n-2) + 1}},
\eea
\end{subequations}
where $\phi_0$ and $C$ are integration constants.
Now, what we need to check is how the above solutions for the scalar field 
and the scale factor are constraining 
our solution for the bulk spacetime. For this
we further specify our bulk potential for the scalar field as
\be
V(\phi) = V_0 e^{\theta \phi}
\ee
where, $V_0$ is constant.
By using Eqs.(\ref{sol}) 
for $R$ and $\phi$ as solutions ansatz
and the bulk potential for the scalar field, 
one obtains different types of solution \cite{debu} which
are characterized by the bulk parameters and suitable boundary
conditions. We will study those solutions 
and their cosmological implication in details else where. 
In this paper we will 
take one particularly simple solution and study its cosmological behaviour.
The solution we are considering is for a simple choice of
parameters $\zeta,\gamma,\theta$ and $\alpha$ setting to zero. We, therefore,
do not have any non-trivial dilaton field in our background. We are also 
interested in the domain wall universe with a spatially flat i.e $k=0$ section.
In our framework, therefore, a spatially flat domain wall is moving in a 
black brane background. One also notes that for the aforementioned 
value of the parameters 
the bulk and brane potential turn into simple
cosmological constant and brane tension respectively. 

Our bulk solution looks like
\bea
N(r) &=& - 2 M r^{-(n-3)} - \left(\frac {2 V_0}{(n-2)(n-1)} - \frac {8 \lambda^2}{(n-2)(n-1)}\right)r^2 +
\frac { {Q^{\dagger}}^2}{(n-3)(n-2)} r^{-2 (n-3)} \nno\\ 
&+& \frac {8 \lambda r^{-(n-4)}} {(n-1)(n-2)} \left( {- \sqrt{4 Q^2 + \lambda^2 r^{2n -4}}}   
+  \frac {4 (n-2) Q^2 r^{-(n-2)}}{\lambda (n-3)}~~ {\cal D}(r,Q) \right) \nno \\
&&R(r) = r ~~~~;~~~ \phi = \phi_0 , 
\eea
where  $M$ and $\phi_0$ are integration constants and
\bea 
{\cal D}(r,Q) = {_2}F_1\left[\frac {n-3} {2n -4}, \frac 1 2 , \frac {3n-7}{2n -4}, -\frac {4 Q^2 r^{-(2n -4)}}
{\lambda^2}\right].
\eea
The solution itself is complicated. For simplicity we study our solution in 
various limits along the radial coordinate and study its behaviour.
If we expand our solution in large r, the expression for the above 
solution becomes
\bea
{N(r)|_{r\rightarrow \infty}} =  - \frac {2 V_0}{(n-2)(n-1)} r^2 
- 2 M r^{-(n - 3)} +
\frac {2 {\tilde Q}^2}{(n-3)(n-2)}r^{-(2n-6)} + {\cal O}(r^{{10 -4n}}),
\eea 
where, ${\tilde Q}^2 = 8 Q^2 + \beta {Q^{\dagger}}^2$.
and for small $r$ limit, 
\bea
 {N(r)|_{r\rightarrow 0}} = \frac {{Q^{\dagger}}^2}
{(n-3)(n-2)} r^{-2 (n-3)} - 2 {\cal M} r^{-(n-3)} 
- \frac {16 \lambda {\cal Q}} {(n-1)(n - 2)} r^{-(n-4)} + {\cal O}(r^2)\nno, 
\eea
where
\bea 
{\cal M} & =& M - \frac {16 Q^2 \Gamma[\frac {3n-7}{2n -4}]\Gamma[\frac {1}{2n-4}]}{\sqrt{\pi}(n-1)(n-3)}
 \left(\frac {4 Q^2}{\lambda^2} \right)^{-\frac {n-3}{2n -4}}\nno\\
{\cal Q} &=& Q\left(1 - \frac { (n-2) \Gamma[\frac {3n-7}{2n -4}]\Gamma[\frac {-1}{2n-4}]}{(n-3) 
\Gamma[\frac {n-3}{2n -4}] \Gamma[\frac {2n -5}{2n -4}]} \right)
\eea
It is, therefore, clear from the above limits that the full complicated
solution for the bulk metric can be cast into the following simple
form  
\bea 
N(r)= \frac {{Q^{\dagger}}^2}{6} r^{-4} - 2 {\cal M} r^{-2}    
- \frac {4 \lambda {\cal Q}} {3} r^{-1} - {\cal H}(r) - \frac {V_0}{6} r^2,
\eea
where, ${\cal H}(r)$ is some complicated function of radial distance $r$. 
But it is important to note that in the 
both limit of $r$ the function is regular i.e.,
\bea
{\cal H}(r)\xrightarrow[]{r\rightarrow 0} {\cal O}(r^2)~~;~~
{\cal H}(r)\xrightarrow[]{r\rightarrow \infty} 2 M r^{-2} -
\frac {{\tilde Q}^2}{3}r^{-4} 
\eea
In the above expressions we considered the number of spacetime 
dimension to be $n=5$ which is of our particular interest.
As one can imagine the expression for ${\cal Q}$ depending 
upon the sign of $Q$ but originally the metric always depends on $Q^2$.
So, for the subsequent discussions, we will take $Q$ to be positive.
The limiting expressions for ${\cal H}(r)$ gives us the
total charge density related to ${\tilde Q}$ and mass density
$M$ of the black hole. The solution has a timelike singularity at $r = 0$.

So far we have discussed about the analytic solution and its
various limiting properties of our bulk spacetime. 
In what follows we will study the dynamics of
a domain wall in that background. 
As is well known \cite{chamblin}, dynamics of a domain wall 
satisfies a Hubble like equation of motion 
\bea
{\dot{R}}^2 + F(R) = 0
\eea
where, "over dot" is the derivative with respect to 
the domain wall proper time $\tau$.
For the simple solution we considered the expression for $F(R)$ tuns out to 
be
\bea
F(R) = N(R) - \frac {{\bar V}_0^2}{36} R^2.
\eea
The form of the potential looks like the   
asymptotic modification of the metric function $N(R)$.

The Hubble equation of motion turns out to be
\bea \label{semi}
H^2 = - \frac {{Q^{\dagger}}^2}{6} R^{-6}  + 2 {\cal M} R^{-4} 
+ \frac {4 \lambda {\cal Q}} {3} R^{-3} + {{\cal H}(R)}{R^{-2}}
+\left(\frac {V_0}{6} + \frac {{\bar V}_0^2}{36}\right)
\eea

As mentioned before the important point we want to emphasize 
here is that effective domain wall equation of motion contains 
a so called "dark matter" energy component in addition to
the usual "dark radiation" term. This is our
new finding which was not discussed in the previous domain
wall study. So the novel feature of our model is that even without
matter field localized on the brane it evolves like a 
standard cosmology. Interestingly this "dark matter"
component is depending upon the charge ($Q$) of the 
Born-Infeld electric field. On the other hand invisible "dark radiation" 
energy depends upon the linear combination of both mass ($M$) and 
Born-Infeld electric charge ($Q$) of the bulk black hole spacetime. 
So the evolution of the domain wall in the Born-Infeld background 
mimics the evolution of the standard cosmology. This is the reason
we call our domain wall dynamics as semi-realistic in nature.

\begin{figure}
\begin{center}
\includegraphics[width=5.50in,height=1.70in]{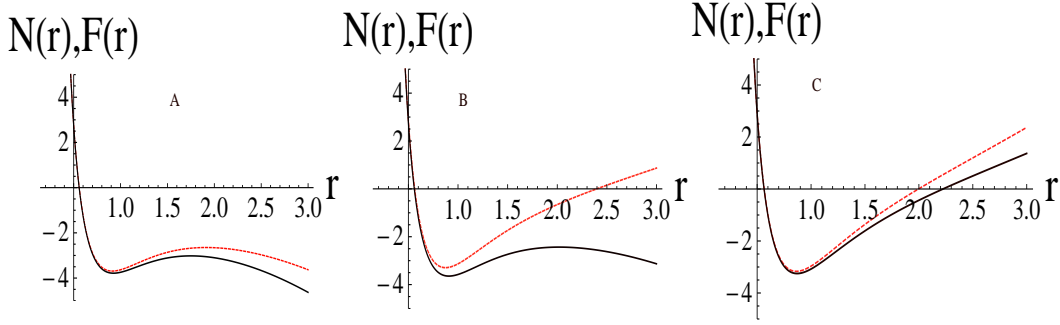}
\end{center}
\caption{Plot for metric function N(r)(dotted red line) and potential F(r)
(solid black line). For these particular plots we set ${Q}^{\dagger}=1, Q=1,
M=3 , \lambda=2 $ with plot (A) $ V_0=2, {\bar V}_0=2$, plot (B) $ V_0 = -1,
{\bar V}_0 = 4$
and plot (C) $V_0 = -2, {\bar V}_0  = -2$.} \label{one}
\end{figure}

In addition to this semi-realistic evolution we also want
to have a bounce in the domain wall dynamics at a 
finite value of its scale factor. This can be achieved by introducing
a Maxwell gauge field in the bulk regarding which we have particularly 
emphasized in the introduction.
From the above Eq.\ref{semi} we see that due to the presence of 
negative energy component so called "stiff matter" induced from
the charge of bulk Maxwell field, we have a bounce followed by
a standard cosmological evolution. Since it is very difficult to get 
an analytic expression from above equation of motion. We plotted the
potential $F(R)$ for different value of the parameters of the model
in Fig.\ref{one} comparing with the bulk $G_{00} = N(R)$ metric component. 
It is clear from the plots that there exists a 
minimum value of the scale factor at which bounce occurs for
three different cases. For the limiting case, near the bouncing point
we can solve the above Hubble equation with the approximation that
the domain wall dynamics is governed by the "stiff matter" and the
"dark radiation". In that limit the solution for the scale factor 
looks like \cite{sudipto}
\bea
R(\eta) = \sqrt{\frac 1 {12 {\cal M}}({{Q}^{\dagger}}^2 + 
24 {\cal M}^2 \eta^2)},
\eea
where, for convenience, we use conformal time $ d\tau = R(\eta) d\eta$
in the above expression. It is clear from the above solution that we 
have minimum value of the scale factor 

\[
R(\eta)_{min} =\sqrt{\frac {{{Q}^{\dagger}}^2} {12 {\cal M}}} ~~~~~~
\mbox{where}~~~~ {\cal M} =  M - \frac {2.5}{\sqrt{\pi}}
\left({4 \lambda Q^2}\right)^{\frac {2}{3}}
\].

In order to have a real solution, mass (M) and BI-charge (Q) of the
black hole should satisfy $M > \frac {2.5}{\sqrt{\pi}}
\left({4 \lambda Q^2}\right)^{\frac {2}{3}}$. As expected 
in the late time evolution is radiation dominated, 
$R(\tau) \sim \tau^{\frac 1 2}$. 

\begin{figure}
\begin{center}
\includegraphics[width=2.60in,height=1.70in]{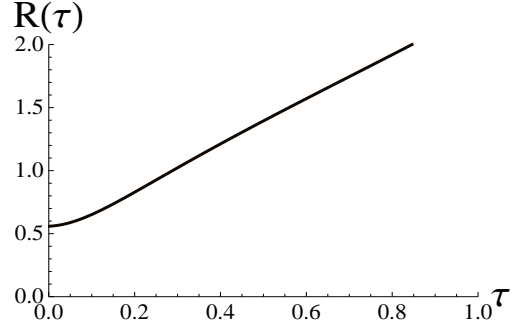}
\end{center}
\caption{Numerical Plot for scale factor $R(\tau)$ for the model $A$ of 
Fig.\ref{one}. So, parameters are ${Q}^{\dagger}=1, Q=1,
M=3 , \lambda=2 $,  $ V_0=2, {\bar V}_0=2$. The minimum value
of the scale factor $R(\tau)_{min}
 = 0.558303 $ } \label{two}
\end{figure}

In order for the completeness 
we also solve the above Hubble equation numerically as shown in Fig.\ref{two}. 
The qualitative feature of the scale factor is same for the different 
parameter values at the bouncing point. So we only plotted the 
scale factor for the model $A$ of Fig.\ref{one}. As Fig.\ref{one} shows 
the effective potential of the domain wall has a minimum which 
leads to an exponential expansion phase of the domain wall after the
bounce. We, therefore, have a natural inflationary phase 
after the bounce but for
a very short period of time. Near the  minimum of the potential 
the scale factor evolves like
\[
R(\tau) \sim e^{\frac {\sqrt{-F''(R_0)} \tau}{2}},
\]
where $F(R)$ has a minimum at $R_0$.

At this point we want
to emphasize that for the two horizon bulk black hole background,
the bounce generically happens inside the inner horizon. Stability
issue on this kind of bounce inside the Cauchy horizon 
has been raised in \cite{myers}. 
Although we think this issue needs further study to completely rule 
out this kind of bouncing cosmological models.
But the general argument says that inner horizon of a charged black
is intrinsically unstable under small perturbation. This instability
is related to the strong cosmic censorship conjecture of a black hole
spacetime. However we are not going to study this issue here any further. 
The point we want to emphasize in our study is that  
for a wide range of parameters of our solution we have charged
under BI gauge field black hole which has no inner horizon. So, for 
those cases stability issue is still not clearly understood. We defer
it for our future study. 


\section{Domain wall cosmology with brane matter field} \label{sec3}

So far we have discussed the case where
there is no realistic matter field localized on the brane.
In this section we will study more realistic situation where
we have radiation as well as normal baryonic matter 
field localized on the domain wall. As we have discussed 
earlier, the dark matter component is induced on the domain
wall through bulk field.
So, the modified Hubble equation for the domain wall turns out
to be
\bea \label{real}
H^2 &=& - \frac {{Q^{\dagger}}^2}{6 R^6}  + \frac {2 {\cal M}} {R^4}
+ \frac {4 \lambda {\cal Q}} {3R^3} +\frac {{\cal H}(R)}{R^2}
+\left(\frac {V_0}{6} + \frac {({\bar V}_0 + 
\rho_{rad} R^{-4} + \rho_{m} R^{-3})^2}{36}\right) \nno \\
&=& \frac 1 {36} \left(\frac {\rho_{rad}^2}{R^8} +
 \frac {2 \rho_{rad}\rho_m}{R^7} + \frac {\rho_m^2}{R^6}\right)- 
\frac {6{Q^{\dagger}}^2}{36 R^6}
 + \frac {36 {\cal M} + {\bar V}_0 \rho_{rad}} {18 R^4}
+ \frac {24 \lambda {\cal Q} +  \bar{V}_0 \rho_m } {18 R^3}\nno\\ 
&& \hskip7.6cm + \frac {{\cal H}(R)}{R^2}
+\left(\frac {V_0}{6} + \frac {{\bar V}_0^2}{36}\right)
\eea
As we can easily identify from the third and fourth terms of the second 
line of the above equation, the induced "dark radiation" ($\rho_{drad}$) 
and the dark matter ($\rho_{dm}$) component on our domain wall 
can be read off as 
\bea
\rho_{drad} = \frac {36 {\cal M}}{{\bar V}_0} = 
\frac {36 }{{\bar V}_0} \left[ M - \frac {2.5}{\sqrt{\pi}}
\left({4 \lambda Q^2}\right)^{\frac {2}{3}}\right]
~~~;~~~\rho_{dm}= \frac {24 \lambda {\cal Q}} {{\bar V}_0} = \frac {96 \lambda Q} {{\bar V}_0}
\eea
with the standard normalization for the Hubble equation 
${\bar V}_0 = \frac {48 \pi}{M_p^2}$, where $M_p$ is the four
dimensional Planck constant.

Now BBN in standard cosmological evolution during 
radiation dominated era as well as the anisotropy 
in the CMB spectrum \cite{subir} tell us that any non-standard radiation like
energy density must be very tiny in order to satisfy the
observed relic abundance. So, the induced "dark radiation" energy 
($\rho_{drad}$) should be much smaller than that of the usual 
radiation density ($\rho_{rad}$). As we mentioned before and
also clear form the above expression for the dark radiation that by
suitably choosing the mass ($M$) and Born-Infield charge ($Q$) of the 
bulk black hole, we can make it zero or very tiny. Furthermore, we 
know that about $23\%$ of the total energy component in our universe
is non-baryonic dark matter in nature. With this  
consideration we can fix the Born-Infeld charge of the black hole
to let say $Q=Q_{dm}$. This observation also fixes the
mass of the bulk black hole to be 
\bea
M \simeq \frac {2.5}{\sqrt{\pi}}
\left({4 \lambda Q_{dm}^2}\right)^{\frac {2}{3}}.  
\eea
Considering the above mass of the black hole, 
we can ignore the "dark radiation" term
in the effective Hubble equation in our subsequent discussions.
At this point it is important point to note about the recent
interests on the additional dark radiation component 
in standard model of cosmology.
There has been a recent speculation that in order to fit some
cosmological observation such as WMAP, the effective number of 
relativistic degrees of freedom in our universe
has to be larger than four \cite{hamann,riess}. 
Even though there is an active debate going on along the line of this
subject. In order to confirm this we need to wait for further 
precision observation such as PLANCK.
Different particle physics model has already been considered 
in order to explain this extra dark radiation component. 
Extra relativistic particle such as sterile neutrino has been introduced as  
a dark radiation\cite{hamann}. Interestingly brane world cosmological
models naturally predict an effective dark radiation component induced 
form the bulk gravitation \cite{deffayet} 
as we also have seen in our current analysis. So by imposing
the constraint coming from this extra cosmological
dark radiation component in our model we can in principle
give precise constraint on the bulk black hole charges. 
For our current study we will consider the dark radiation component 
to be negligibly small.  

Now, further constraint on the black hole parameters will come 
from the bounce for a particular value of the scale factor. 
Since we are considering the case, where the domain wall is very
close to the bouncing point, we can ignore the matter and 
cosmological constant part from the the Hubble Eq.\ref{real} 
and set it to zero right at the bounce.
At the bouncing point we approximated the Eq.\ref{real} to be
\bea
\left(\frac {\rho_{rad}^2}{R_b^8} +
\frac {2 \rho_{rad}\rho_m}{R_b^7} + \frac {\rho_m^2}{R_b^6}\right)-
\frac {6{Q^{\dagger}}^2}{R_b^6} + 
\frac {2 {\bar V}_0 \rho_{rad}} {R_b^4} = 0.
\eea
it is very difficult to get an analytical expression for the scale
factor. As we have checked if the condition below is satisfied then 
we can have a bounce at $R_b$ satisfying the above equation.
\bea
{\rho_{rad}^2}+
{2 \rho_{rad}\rho_m}{R_0} -
({6{Q^{\dagger}}^2}- \rho_m^2) {R_0^2} +  
{2 {\bar V}_0 \rho_{rad}} {R_0^4} \leq 0,
\eea
where

\[
R_0=  \frac { 2\cdot 3^{\frac 1 3}  a b + 2^{\frac 1 3}  
\left (-9 a^2  c +  \sqrt{3} \sqrt{a^3 
 (-4 b^3  + 27 a c^2 )}\right)^{\frac 2 3}} 
{6^{\frac 2 3}a \left(-9 a^2  c +  \sqrt{3} 
\sqrt{a^3  (-4 b^3  + 27 a c^2 )}\right)^{\frac 1 3}}
\] 
with 

\[
a = {8 {\bar V}_0 \rho_{rad}}~~;~~b= 2 ({6{Q^{\dagger}}^2}- \rho_m^2)~~;~~
c= {2 \rho_{rad}\rho_m}
\]

If we consider only the radiation field on the brane then all
the above expressions becomes simple. To simplify the subsequent
analysis, let us consider $\rho_m=0$, then in order to get
a bounce one needs to satisfy \cite{sudipto}
\bea
{Q^{\dagger}}^4 \geq \frac 2 9 {\bar V}_0 \rho_{rad}^2 = 
\frac {32 \pi \rho_{rad}^3} {3 M_p^2}
\eea
 So, the bounce restricts the value
of electro-magnetic charge of the bulk black hole. Then if the
above bound is satisfied, the minimum value for the scale factor 
approximately amounts to
\bea
R_b = R_{min} \simeq \left( \frac {384 \pi 
\rho_{rad}^3}{M_p^2}\right)^{\frac 1 4}  
\eea
In conclusion we have seen that dynamics of a domain wall in the 
Maxwell-Born-Infeld black hole background is semi-realistic in
nature. The dynamics of the domain wall is governed by the
dark radiation, dark matter and cosmological constant all of which
can be induced from the above mentioned static bulk black hole 
charges. In addition to the above semi-realistic cosmological expansion,
we also have seen that our model passes through a bouncing phase
as well. All these interesting features give us a hope that
brane domain wall model could be an interesting framework to construct
a singularity free cosmological models. In the next section we
will discuss about the perturbation across the domain wall
junction.   
\section{Perturbation} \label{sec4}
In this section we will try to set up the stage for the scalar perturbation
in dynamic domain wall scenario for our future study. Detail study on the
perturbation dynamics in the framework of domain wall scenario has not
been studied yet. The present paper is beyond the scope of this study. In this
section will begin this programme by first calculating how the 
perturbed Israel junction condition across the domain wall looks like.
Some part of this calculation can be found in many paper dealing with
perturbation in a brane-world scenario (see the review \cite{roy}).
First we will start with a general form
of the bulk background metric 
\bea
ds^2 = G_{ab} dX^a dX^b = - n(r)^2 dt^2 + b(r)^2 dr^2 + 
R(r)^2 (dx^2 + dx^2 + dx^2)
\eea
where, for our particular case  $n(r) = \frac 1 {b(r)} = N(r)$. 
We parametrize our brane as 
\be
X^{a} = {\bar X}^{a}(y^{\mu})~~\mbox{where}~~{\bar X}^0 =t \equiv t(\tau)~~;
~~{\bar X}^1= r\equiv r(\tau)
\ee
Four tangent vectors to the brane are
\bea
{\bar V}^a_{\mu} = \frac {\pr {\bar X}^{a}}{\pr y^{\mu}} ~~\implies~~
{\bar V}_{\eta}^a = \left(\frac {\pr t}{\pr \tau},0,0,\frac {\pr r }
{\pr \tau}\right)~~;
~~{\bar V}_i^a = (0,\delta_i^a,0)
\eea
The normal vector to the brane would be
\bea
{\bar n}_a = \frac {b n {\dot r}} {\gamma} \delta_a^t + \frac {1} {\gamma}
\delta_a^r,
\eea
where $\gamma = {\sqrt{n^2 - b^2 {\dot r}^2}}$. The normal vector
satisfies  ${\bar V}_{\mu}^{a} {\bar n}_a =0$ and normalizibility
condition $ {\bar n}_a {\bar n}^a =1$.  

The from of the induced brane metic is defined as before
\bea
ds_{brane}^2 = G_{ab} V^a_{\mu}V^b_{\nu} dy^{\mu} dy^{\nu} = - d\tau^2 + 
R(\tau)^2 (dx^2 + dy^2 + dz^2)
\eea
with 
\bea
\left(\frac {dt}{d\tau}\right) = \frac  1 {\sqrt{n^2 - b^2 {\dot r}^2}} =
\frac 1 {\gamma}
\eea  
Now, in this background set up, we will consider the linear 
perturbation with a dynamic domain wall.
We will consider the scalar perturbation in our background. 
The linearly perturbed Einstein equations of motion takes the following form
\bea
\delta R_{AB} = \delta T_{AB}^{\phi} +\delta T_{AB}^{{\cal A}} + 
\delta T_{AB}^{\cal B} ~~~;~~~
\delta K_{MN} = - \frac 1 {2(n -2)} \delta[\bar V(\phi) h_{MN}]
\eea

with the Background metric perturbation 
$g_{ab} = {\bar G}_{ab} + \delta G_{ab}$ as
where,
\bea
\delta G_{ab} = \left(\begin{array}{ccc}
-2 n^2 {\cal W} & R^2 {\cal X}_{,i} & n {\cal W}_r\\
R^2 {\cal X}_{,i} & R^2[2 {\cal S}\delta_{ij} + 2 {\cal E}_{,ij}]& R^2 
{\cal X}_{r,i}\\
n {\cal W}_r & R^2 {\cal X}_{r,j} & 2 b^2 {\cal W}_{rr}
\end{array}\right)
\eea  
As we know in five dimension, the gauge transformation 
$x^a \rightarrow x^a + \xi^a$ has three scalar function. So, by choosing these
three scalar function we set three scalar function in $\delta G_{ab}$ to zero.
After choosing this gauge, the metric perturbation becomes,
\bea
\delta G_{ab} = \left(\begin{array}{ccc}
-2 n^2 {\cal W} & 0 & n {\cal W}_r\\
0 & 0 & R^2 
{\cal X}_{r,i}\\
n {\cal W}_r & R^2 {\cal X}_{r,j} & 2 b^2 {\cal W}_{rr}
\end{array}\right)
\eea 
Finally we, therefore, have four scalar degrees of freedom.
Furthermore, since we have domain wall which breaks the 
translational invariance along the radial direction, we
have brane fluctuating mode. Let us parametrize the perturbed 
brane position as 
$X^{a} = {\bar X}^{a} + \chi^{a}(y^{\mu})$, where $y^{\mu}$ 
is the brane coordinate. 
The fluctuation vector field $\chi^a$ can be conveniently decomposed  
as
\bea
\chi^a = \xi^{\mu} {\bar V}_{\mu}^a + \zeta {\bar n}^a
\eea
where $(\zeta,\xi^{\mu})$ are the five arbitrary function defining 
the fluctuating domain wall coordinate. However, we also
have a re-parametrization 
invariance in domain wall coordinate $y^{\mu}$. So, we can again 
fix this gauge by choosing 
$\xi^{\mu} =0$. The domain wall fluctuation can, therefore, be parametrized
by a single function $\zeta$. 
The perturbed induced metric on the brane therefore would be
\bea
\delta h_{\mu\nu} = \delta G_{ab} {\bar V}_{\mu}^a {\bar V}_{\nu}^b +
2 \zeta {\bar K}_{\mu\nu},
\eea
where, ${\bar K}_{\mu\nu} = {\bar V}_{\mu}^a {\bar V}_{\nu}^b 
{\bar \nabla}_a {\bar n}_b$ is 
the background extrinsic curvature of the brane.

The form of the perturbed normal vector to the domain wall 
$\delta n_a$ takes the following form:
\bea
\delta n_i &=& -\partial_i \zeta \nno\\
\delta n_t &=& \frac {{\bar n}_t} 2 {\cal D}_1 - \frac {n^2}{\gamma^2} 
{\cal D}_2 \nno \\
\delta n_r &=& \frac {{\bar n}_r} 2 {\cal D}_1 - \frac 
{b^2 {\dot r}}{\gamma^2} {\cal D}_2     
\eea

where
\bea
 {\cal D}_1  & =& \delta G_{ab} {\bar n}^a 
{\bar n}^b + \zeta {\bar n}^c \pr_c {\bar G}_{ab}
{\bar n}^a {\bar n}^b~~~;~~~{\cal D}_2= {\dot \zeta} + \zeta {\bar G}_{ab} 
{\bar n}^a {\bar V}^c_{\tau} \gamma \partial_c {\bar n}^b \nno
\eea 
Equipped with all the above relevant variations 
the expression for the perturbed extrinsic curvature becomes
\bea
{\delta K}_{ij} &=& - \zeta_{,ij} + \frac {R R'}{n^2} \delta n_r
 + \Delta \Gamma^a_{ij} {\bar n}_a , \\
{\delta K}_{0i} &=&  \frac 1 2 \frac {\partial_i{\zeta} {\bar n}^b}{\gamma}(\bar{\nabla}_b
\bar{n}_t + \bar{\nabla}_t \bar{n}_b+ \dot{r} \bar{\nabla}_b 
\bar{n}_r + \dot{r} \bar{\nabla}_r \bar{n}_b) + 
\bar{V}^b_{\tau}(
\frac 1 2(\bar{\nabla}_i \delta{n}_b + \bar{\nabla}_b \delta{n}_i) + 
\Delta \Gamma^c_{ib}\bar{n}_c), \\
{\delta K}_{00} &=& \frac {\dot{\zeta} {\bar n}^b}{\gamma^2}(\bar{\nabla}_b
\bar{n}_t + \bar{\nabla}_t \bar{n}_b+ \dot{r} \bar{\nabla}_b
\bar{n}_r + \dot{r} \bar{\nabla}_r \bar{n}_b) + \bar{V}^a_{\tau} \bar{V}^b_{\tau}(
\frac 1 2(\bar{\nabla}_a \delta{n}_b + \bar{\nabla}_b \delta{n}_a) +
\Delta \Gamma^c_{ab}\bar{n}_c) 
\eea
where,
\bea
\Delta \Gamma^c_{ab} &=& \delta \Gamma^c_{ab} + \zeta 
{\bar n}^r \pr_r{{\bar \Gamma}^c_{ab}} \nno\\
\eea 
In this section we have computed the perturbing boundary condition across 
the junction of dynamic domain wall. We will
do the detailed analysis of this perturbation in our subsequent paper.

\section{Conclusion}\label{con}
Standard model of cosmology is one of the most successful 
models in successfully explaining the evolution of our 
universe. There are some important fundamental issues
in this model which have been puzzling physicists for a long time.
As we have been mentioning through out our present paper that
our universe under the standard model of cosmology encountered a 
singularity as we go backward in time.
This is definitely unexpected for any physically meaningful
theory. There has been lot of attempts to construct effective models
which can avoid this big-bang singularity. As we have mentioned,
higher dimensional cosmological model has particularly gained considerable 
interest in this respect. In this paper we have studied dynamic
domain wall cosmology where we can realise the bouncing cosmology.
People have already found this kind of bouncing solution before
\cite{sudipto}, but the interesting finding in our model is the possibility
of inducing "dark matter" like energy component on the the domain wall
by considering a simple well known fields in the bulk. 
This aspect leads us to construct a semi-realistic 
bouncing domain wall cosmology by introducing different types of 
gauge field in the higher dimensional background. 
We have considered Maxwell-Born-Infeld gauge field
background in the bulk and studied the dynamics of 
the domain wall in those background. 
We found out the analytic bulk spacetime solutions taking into account the 
back-reaction of those gauge fields and the dynamic domain walls.
There exits many different types of solutions depending upon  
various choices of parameters \cite{debu}. 
In this paper we discussed about a particularly simple solution in which
the dynamics of the domain wall mimics a semi-realistic cosmological
evolution along the extra dimension compared to our 
standard cosmological scenario.  

As we already mentioned the important aspects of our model is the presence
of "dark matter" like energy component which is induced from the 
bulk Born-Infeld charge. In addition to this we have standard 
"dark radiation" component coming from the black hole mass (M) and 
Born-Infeld charge (Q). 
In addition to the standard evolution an effective negative energy 
density is induced from the bulk usual electromagnetic charge $(Q^{\dagger})$
leading to a singularity free bounce of the domain wall at finite value of 
its scale factor.  
All these aspects provide us an interesting possibility to 
construct a realistic bouncing domain wall cosmology. Furthermore it gives us
a hint that may be the domain wall framework could be an interesting 
play ground to solve long standing dark matter and dark energy 
problem in our universe. Perturbation analysis in this kind of model
is very important in regard to the stability of itself as well as
the CMB observation. We have just initiated this in our current
paper which shows fairly complicated set of equations only for the
perturbed junction condition across the domain wall. In our next paper we
will consider this in detail.


\end{document}